\input harvmac.tex

\newread\epsffilein    % file to \read
\newif\ifepsffileok    % continue looking for the bounding box?
\newif\ifepsfbbfound   % success?
\newif\ifepsfverbose   % report what you're making?
\newdimen\epsfxsize    % horizontal size after scaling
\newdimen\epsfysize    % vertical size after scaling
\newdimen\epsftsize    % horizontal size before scaling
\newdimen\epsfrsize    % vertical size before scaling
\newdimen\epsftmp      % register for arithmetic manipulation
\newdimen\pspoints     % conversion factor
\pspoints=1bp          % Adobe points are `big'
\epsfxsize=0pt         % Default value, means `use natural size'
\epsfysize=0pt         % ditto
\def\epsfbox#1{\global\def\epsfllx{72}\global\def\epsflly{72}%
   \global\def\epsfurx{540}\global\def\epsfury{720}%
   \def\lbracket{[}\def\testit{#1}\ifx\testit\lbracket
   \let\next=\epsfgetlitbb\else\let\next=\epsfnormal\fi\next{#1}}%
\def\epsfgetlitbb#1#2 #3 #4 #5]#6{\epsfgrab #2 #3 #4 #5 .\\%
   \epsfsetgraph{#6}}%
\def\epsfnormal#1{\epsfgetbb{#1}\epsfsetgraph{#1}}%
\def\epsfgetbb#1{%
%
%   The first thing we need to do is to open the
%   PostScript file, if possible.
%
\openin\epsffilein=#1
\ifeof\epsffilein\errmessage{I couldn't open #1, will ignore it}\else
%
%   Okay, we got it. Now we'll scan lines until we find one that doesn't
%   start with %. We're looking for the bounding box comment.
%
   {\epsffileoktrue \chardef\other=12
    \def\do##1{\catcode`##1=\other}\dospecials \catcode`\ =10
    \loop
       \read\epsffilein to \epsffileline
       \ifeof\epsffilein\epsffileokfalse\else
%
%   We check to see if the first character is a % sign;
%   if not, we stop reading (unless the line was entirely blank);
%   if so, we look further and stop only if the line begins with
%   `%%BoundingBox:'.
%
          \expandafter\epsfaux\epsffileline:. \\%
       \fi
   \ifepsffileok\repeat
   \ifepsfbbfound\else
    \ifepsfverbose\message{No bounding box comment in #1; using defaults}\fi\fi
   }\closein\epsffilein\fi}%
%
%   Now we have to calculate the scale and offset values to use.
%   First we compute the natural sizes.
%
\def\epsfclipstring{}% do we clip or not?  If so,
\def\epsfsetgraph#1{%
   \epsfrsize=\epsfury\pspoints
   \advance\epsfrsize by-\epsflly\pspoints
   \epsftsize=\epsfurx\pspoints
   \advance\epsftsize by-\epsfllx\pspoints
%
%   If `epsfxsize' is 0, we default to the natural size of the picture.
%   Otherwise we scale the graph to be \epsfxsize wide.
%
   \epsfxsize\epsfsize\epsftsize\epsfrsize
   \ifnum\epsfxsize=0 \ifnum\epsfysize=0
      \epsfxsize=\epsftsize \epsfysize=\epsfrsize
      \epsfrsize=0pt
%
%   We have a sticky problem here:  TeX doesn't do floating point arithmetic!
%   Our goal is to compute y = rx/t. The following loop does this reasonably
%   fast, with an error of at most about 16 sp (about 1/4000 pt).
% 
     \else\epsftmp=\epsftsize \divide\epsftmp\epsfrsize
       \epsfxsize=\epsfysize \multiply\epsfxsize\epsftmp
       \multiply\epsftmp\epsfrsize \advance\epsftsize-\epsftmp
       \epsftmp=\epsfysize
       \loop \advance\epsftsize\epsftsize \divide\epsftmp 2
       \ifnum\epsftmp>0
          \ifnum\epsftsize<\epsfrsize\else
             \advance\epsftsize-\epsfrsize \advance\epsfxsize\epsftmp \fi
       \repeat
       \epsfrsize=0pt
     \fi
   \else \ifnum\epsfysize=0
     \epsftmp=\epsfrsize \divide\epsftmp\epsftsize
     \epsfysize=\epsfxsize \multiply\epsfysize\epsftmp   
     \multiply\epsftmp\epsftsize \advance\epsfrsize-\epsftmp
     \epsftmp=\epsfxsize
     \loop \advance\epsfrsize\epsfrsize \divide\epsftmp 2
     \ifnum\epsftmp>0
        \ifnum\epsfrsize<\epsftsize\else
           \advance\epsfrsize-\epsftsize \advance\epsfysize\epsftmp \fi
     \repeat
     \epsfrsize=0pt
    \else
     \epsfrsize=\epsfysize
    \fi
   \fi
%
%  Finally, we make the vbox and stick in a \special that dvips can parse.
%
   \ifepsfverbose\message{#1: width=\the\epsfxsize, height=\the\epsfysize}\fi
   \epsftmp=10\epsfxsize \divide\epsftmp\pspoints
   \vbox to\epsfysize{\vfil\hbox to\epsfxsize{%
      \ifnum\epsfrsize=0\relax
        \includegraphics{#1}%
      \else
        \epsfrsize=10\epsfysize \divide\epsfrsize\pspoints
        \includegraphics{#1}%
      \fi
      \hfil}}%
\global\epsfxsize=0pt\global\epsfysize=0pt}%
%
%   We still need to define the tricky \epsfaux macro. This requires
%   a couple of magic constants for comparison purposes.
%
{\catcode`\%=12 \global\let\epsfpercent=%\global\def\epsfbblit{%BoundingBox}}%
%
%   So we're ready to check for `%BoundingBox:' and to grab the
%   values if they are found.
%
\long\def\epsfaux#1#2:#3\\{\ifx#1\epsfpercent
   \def\testit{#2}\ifx\testit\epsfbblit
      \epsfgrab #3 . . . \\%
      \epsffileokfalse
      \global\epsfbbfoundtrue
   \fi\else\ifx#1\par\else\epsffileokfalse\fi\fi}%
%
%   Here we grab the values and stuff them in the appropriate definitions.
%
\def\epsfempty{}%
\def\epsfgrab #1 #2 #3 #4 #5\\{%
\global\def\epsfllx{#1}\ifx\epsfllx\epsfempty
      \epsfgrab #2 #3 #4 #5 .\\\else
   \global\def\epsflly{#2}%
   \global\def\epsfurx{#3}\global\def\epsfury{#4}\fi}%
%
%   We default the epsfsize macro.
%
\def\epsfsize#1#2{\epsfxsize}
%
%   Finally, another definition for compatibility with older macros.
%

%\input epsf
%
\ifx\epsfbox\UnDeFiNeD\message{(NO epsf.tex, FIGURES WILL BE
IGNORED)}
\def\Fig.in#1{\vskip2in}% blank space instead
\else\message{(FIGURES WILL BE INCLUDED)}\def\Fig.in#1{#1}\fi
\def\iFig.#1#2#3{\xdef#1{Fig..~\the\Fig.no}
\goodbreak\topinsert\Fig.in{\centerline{#3}}%
\smallskip\centerline{\vbox{\baselineskip12pt
\advance\hsize by -1truein\noindent{\bf Fig..~\the\Fig.no:} #2}}
\bigskip\endinsert\global\advance\Fig.no by1}
%%%%%%%%%%%%%%%%%%%%%%%%%%%%%%%%%%%%%%%%%%%%%%%%%%%%%%%%%%%%%%%%

\def\cst {{\rm const.}}

\def \ov {\over}

\def \lr { \lref}

\def\dd {\partial }

\def\l{\lambda}

\def\n{\noindent}
\gdef \jnl#1, #2, #3, 1#4#5#6{ { #1~}{ #2} (1#4#5#6) #3}
\def\np {  Nucl.  Phys. }
\def \pl { Phys. Lett. }
\def \mpl { Mod. Phys. Lett. }
\def \prl { Phys. Rev. Lett. }
\def \pr  { Phys. Rev. }
\def \cqg { Class. Quant. Grav. }
\def \jmp { J. Math. Phys. }

\def \grg {Gen. Rel. Grav. }

\lr \bilcal {A. Bilal and C. Callan, \np B394 (1993) 73;
 S. de Alwis, \pl B289 (1992) 278; B300 (1993) 330.}
\lr \cghs {C. Callan, S. Giddings, J. Harvey and A. Strominger,
\pr D45 (1992) R1005. }
\lr \witt {E. Witten, \pr D44 (1991) 314. }
\lr \birdav {N.D. Birrell and P.C.W. Davies, {\it Quantum fields in
curved space} (Cambridge University Press, Cambridge, England, 1982). }
\lr \schutz {B.F. Schutz, `` A first course in general relativity",
(Cambridge University Press, Cambridge, England, 1985).}
\lr \chrisful {S.M. Christensen and S.A. Fulling, \pr D15 (1977) 2088. }
\lr \gidnel {S.B. Giddings and W.M. Nelson, \pr D46 (1992) 2486. }
\lr \hawk {S.W. Hawking, Commun. Math. Phys. 43 (1975) 199. }
\lr \polyakov {A.M. Polyakov, \pl 163B (1981) 207. }
\lr \rustse{H. Verlinde, in
Proceedings of the School in ``String theory and quantum gravity",
Trieste, ICTP (ed. J. Harvey et al) (1991) ;
J.G. Russo and A.A. Tseytlin, \np B382 (1992) 259; A. Strominger,
\pr D46 (1992) 4396.}
\lr \rst {J.G. Russo, L. Susskind and L. Thorlacius,
\pr D46 (1992) 3444; \pr D47 (1993) 533.  }

\lr\horo {G. Horowitz, in Proceedings of the School in ``String theory and
quantum gravity", Trieste, 1992. }

\lr\melv { M.A. Melvin, Phys. Lett. 8 (1964) 65.}

\lr\myers{Y. Kazama, Y. Satoh and A. Tsuchiya,
\pr D51 (1995) 4265;
G. Michaud and  R.C.  Myers, ``Two-dimensional dilaton black holes",
gr-qc/9508063. }

\lr\mazzi {D. Mazzitelli and J.G. Russo, \pr D47 (1993) 4490.  }

\lr\harhaw{J.B. Hartle and S.W. Hawking, \pr D28 (1983) 2960.}

\lr\gibb {G.H. Gibbons and K. Maeda, \np B298 (1988) 741.}

\lr\tsey {A.A. Tseytlin, ``Exact solutions of closed string theory",
 hep-th/9505052.}

\lr\das{ S.R. Das and S. Mikherji,  \mpl A9 (1994)  3105 ; T. Chung and
H. Verlinde,  \np B418 (1994) 305.}

\lr \beken {J.D. Bekenstein, \pr D7 (1973) 2333; D9 (1974) 3292.  }
\lr \hawk {S. Hawking, Commun. Math. Phys. 43 (1975) 199.  }
\lr \thoof { G. 't Hooft,  \np B256 (1985) 727.  }
\lr \hooft { G. 't Hooft, Physica Scripta T36 (1991) 247;
 {\it Dimensional reduction in quantum gravity},
Utrecht preprint THU-93/26, gr-qc/9310006. }
\lr \sussk {L. Susskind,  {\it The world as a hologram},
  preprint SU-ITP-94-33, hep-th/9409089. }
\lr \susugl {L. Susskind and J. Uglum, \pr D50 (1994) 2700.}
\lr \suss {L. Susskind,  {\it Some speculations about entropy in
string theory},    RU-93-44, hep-th/9309145;
J.G. Russo and L. Susskind, \np B 437 (1995) 611; A.~Sen,
{\it Extremal black holes and elementary string states},
  TIFR-TH-95-19, hep-th/9504147; A. Peet, {\it Entropy and supersymmetry of $D$ dimensional extremal electric black holes versus string states},
  PUPT-1548, hep-th/9506200.}
\lr \sredn {M. Srednicki, \prl 71 (1993) 666;
V. Frolov and I. Novikov, \pr D48 (1993) 4545;
D. Kabat and M.J. Strassler, \pl B329 (1994) 46;
C. Callan and F. Wilczeck; \pl B333 (1994) 55.}
\lr \stu {L. Susskind, L. Thorlacius and J. Uglum,
\pr D48 (1993) 3743.  }
\lr \membr {K. Thorne, R. Price and D. MacDonald, {\it Black holes:
the membrane paradigm} (Yale Univ. Press, New Haven, CT, 1986). }
\lr \entropy {E. Keski-Vakkuri and S. Mathur, \pr D50 (1994) 917;
T. Fiola, J. Preskill, A. Strominger
and S. Trivedi, \pr D50 (1994) 3987;  R.C. Myers, \pr D50 (1994) 6412;
J.D. Hayward, DAMTP-R94-61, gr-qc/9412065.
}
\lr \rstf  {J.G. Russo, L. Susskind and L. Thorlacius, \pl B292 (1992) 13.}
\lr \veil  {J.G. Russo, \pr D49 (1994) 5266.}
\lr\wald {R. M. Wald, {\it General Relativity} (University of
Chicago Press, Chicago, 1984).}
 \lr \svv {K. Schoutens, H. Verlinde and E. Verlinde, {\it
Black hole evaporation and quantum gravity},
CERN-TH.7142/94.}
\lr \kant{R. Laflamme and E.P. Shellard, \pr D35 (1987) 2315;
J. Louko, \it{ibid.} 35 (1987) 3760;
J. Louko and T. Vachaspati, \pl B223 (1989) 21;
A. Gangui, F.D. Mazzitelli and M. Castagnino, \pr D43 (1991) 1853.}
\lr\ksacks{R. Kantowski and R. Sacks,\jmp 7 (1967) 443.}
\lr\wein{S. Weinberg, {\it Gravitation and Cosmology},
John Wiley, Inc., New York (1972).}
\lr\cosmol{ M. Hotta, Y. Suzuki, Y. Tamiya and M. Yoshimura,
Progr. Th. Phys. 90 (1993) 689; G. Martin and F.D. Mazzitelli, \pr D50 (1994) 613;
J.S.F. Chan and R.B.~Mann, \cqg 12 (1995) 351;
J. Lidsey, \pr D51 (1995) 6829.}
\lr\faru{A. Fabbri and J.G. Russo, {\it Soluble models in 2d dilaton gravity},
hep-th/9510109 (to appear in \pr D).}
\lr\mann{J.D. Brown, M. Henneaux and C. Teitelboim, \pr D33 (1986), 319;
R.B. Mann, \grg 24 (1992), 433.} 
\lr\klostr{T. Kl\"osch and T. Strobl, {\it Classical and Quantum Gravity
in 1+1 dimensions Part I: A unifying approach}, gr-qc/9508020;
{\it Classical and Quantum Gravity in 1+1 dimensions Part II: The universal
coverings}, gr-qc/9511081.}
\lr\schmidt{H.J. Schmidt, \jmp 32 (1991) 1562; S. Mignemi and H.J. Schmidt,
\cqg 12 (1995) 849.} 
\lr\trivedi{S.P. Trivedi, \pr D47 (1993) 4233.}
\lr\mignemi{M. Cadoni and S. Mignemi, {\it On the conformal equivalence between
2d black holes and Rindler spacetime}, gr-qc/9505032.}
\lr\leka{J.S. Lemos and P.M. Sa, \pr D49 (1994) 2897; M.O. Katanaev,
W. Kummer and H. Liebl, {\it Geometric interpretation and classification of
global solutions in generalized dilaton gravity}, gr-qc/9511009.}

\baselineskip8pt
\Title{\vbox
{\baselineskip 6pt 
\hbox{SISSA-ISAS/24/96/EP} 
{\hbox{gr-qc/9602047}}
{\hbox{
   }}} }
{\vbox{\centerline { Two-dimensional black holes in accelerated frames:} 
\vskip .2in \centerline {Spacetime structure}
}}
\bigskip\bigskip\bigskip
\vskip -20 true pt
\centerline  { {R. Balbinot  }}
 \smallskip \bigskip
\centerline{\it  Dipartimento di Fisica dell'Universit\`a di Bologna and
INFN sezione di Bologna,}
\smallskip
\centerline{\it  Via Irnerio 46, 40126 Bologna, Italy}
\smallskip
\centerline {\it   balbinot@bologna.infn.it}
\bigskip\bigskip\bigskip
\vskip -20 true pt
\centerline { A. Fabbri }
\smallskip \bigskip
\centerline {\it SISSA-ISAS and INFN sezione di Trieste,  }
\smallskip
\centerline {\it Via Beirut 2-4, 34014 Trieste, Italy}
\smallskip
\centerline {\it   fabbri@gandalf.sissa.it}
\bigskip\bigskip\bigskip
\bigskip\bigskip\bigskip
\centerline {\bf Abstract}
\bigskip

Analytical extensions and resulting Penrose diagrams are given for the 
solutions of a simple 2d dilaton gravity theory describing 
black holes which are static as viewed by asymptotically 
accelerated observers.

\medskip
\baselineskip8pt
\noindent

\Date {February 1996}

%\draftmode
\noblackbox
\baselineskip 14pt plus 2pt minus 2pt
%\baselineskip 20pt plus 2pt minus 2pt
%%%%%%%%%%%%%%%%%%%%%%%%%%%%%%%%%%%%%%%%%%%%%%%%%%%%%%%%%%%%%%%%

\vfill\eject

%\draftmode
%%%%%%%%%%%%%%%%%%%%%%%%%%%%%%%%%%%%%%%%%%%%%%%%%%%%%%%%%%%%%%%%
\newsec{ Introduction}
%%%%%%%%%%%%%%%%%%%%%%%%%%%%%%%%%%%%%%%%%%%%%%%%%%%%%%%%
Two-dimensional theories of gravity represent very simple schemes
used nowadays to investigate issues related to black holes evaporation.
\par \n 
One of the most popular models, the RST \rst, has the advantage of being
exactly solvable; it describes in a closed analytical form the
semiclassical physics of an evaporating black hole (i.e. formation,
evaporation and backreaction).
\par
In a previous paper \faru\ it has been shown 
that there exists a one parameter family of
classical theories all leading to the RST action at the semiclassical
level. These theories are described by the action
\eqn\accla{
S_n={1\ov{2\pi}}\int d^2 x \sqrt{-g}[ e^{-{2\ov n}\phi}
(R + {4\ov n}(\nabla\phi)^2) + 4\l^2 e^{-2\phi}],
}
where $R$ is the scalar curvature associated to the two-dimensional
metric tensor $g_{ab}$, $\phi$ is the dilaton field.
In the case $n=1$, eq. \accla\ is the usual CGHS action 
\cghs.
\par \n
The geometries which extremize $S_n$ have been shown to have typically
a black hole structure with curvature singularities, horizons and regions
where $R$ vanishes.
\par \n
However, their most striking feature is that the natural frame in which 
the metric is static is not `asymptotically' minkowskian 
but Rindler like.\foot{ This seems a common feature of 2d black holes,
see for example \mann. }
Depending on the value of the parameter $n$, the `asymptotic region' 
where the spacetime becomes flat ($R\to 0$) corresponds either to true
infinity (in the sense that geodesic observers take an infinite amount
of their proper time to reach this region) or just to the location of another
horizon (besides the black hole one), the acceleration horizon, which
inertial observers can cross in a finite proper time.
\par
All the solutions possess a Killing vector whose norm vanishes on one
or more horizons. The analytical extension beyond these horizons
reveals a very rich and interesting space-time structure which is the
object of the present paper.\foot{
 The spacetime structure of other models of 2d black holes are discussed
 in Refs \leka\ and \klostr.}
%%%%%%%%%%%%%%%%%%%%%%%%%%%%%%%%%%%%%%%%%%%%%%%%

\newsec{ The theory and its solutions}

%%%%%%%%%%%%%%%%%%%%%%%%%%%%%%%%%%%%%%%%%%%%%%%%%

For the theory described by the action $S_n$, 
the equations of motion are
\eqn\eqmo{
g_{\mu\nu}\big[ {4\ov n}\big( -{1\ov 2}+{1\ov n}\big) (\nabla \phi)^2 -{2\ov n}
\nabla^2 \phi - 2\lambda^2 e^{{{2-2n}\ov n}\phi}\big] +
{4\ov n}\big( 1-{1\ov n}\big) \dd_{\mu}\phi \dd_\nu \phi
}
$$
+{2\ov n}
\nabla_{\mu}\dd_{\nu}\phi =0 \ ,
$$
\eqn\dquattro{
{R\ov n} - {4\ov {n^2}}(\nabla \phi)^2
+{4\ov n}\nabla^2\phi +4
\lambda^2 e^{{{2-2n}\ov {n}}\phi}=0\ .
}
The solution of these equations is easily derived in the conformal gauge
$g_{\pm\pm}=0$, 
\par\n
$g_{+-}=-{1\ov 2}e^{2\rho}$.
Eqs \eqmo\ and \dquattro\ then become
\eqn\confg{
-{4\ov {n^2}}\dd_+\phi\dd_-\phi+ {2\ov n}\dd_+\dd_-\phi
- \lambda^2 e^{{{2-2n}\ov n}\phi + 2\rho} =0\ ,
}
\eqn\dnove{
{2\ov n} \dd_+\dd_-\rho +{4\ov{n^2}}\dd_+\phi\dd_-\phi
-{4\ov n}\dd_+\dd_-\phi +\lambda^2 e^{{{2-2n}\ov n}\phi+
2\rho} =0\ .
}
These equations have to be supplemented by the constraints 
\eqn\const{
e^{-{2\ov n}\phi}
\big[ {4\ov n}\big( 1-{1\ov n}\big) \dd_{\pm}\phi\dd_{\pm}\phi +
{2\ov n}\dd_{\pm}^2\phi -{4\ov n}\dd_{\pm}\rho \dd_{\pm}\phi \big]
 =0\ .
}
From \confg\ and \dnove\ it follows that 
\eqn\gauge{
{2\ov n} \dd_+\dd_- (\rho - \phi)=0,
}
which implies
\eqn\cio{
\rho = \phi + f_+(x^+) + f_-(x^-).
}
One can now perform a coordinate transformation $x^{\pm}\to x^{'\pm}=
F_{\pm}(x^{\pm})$ which preserves the conformal gauge and for which
$\rho=\phi$.
\par \n
In this `Kruskal' gauge the remaining field equation takes the form
\eqn\kru{
\dd_+\dd_-(e^{-{2\ov n}\phi})=-\l^2
}
and the constraints 
\eqn\cici{
\dd_{\pm}^2 (e^{-{2\ov n}\phi})=0.
}
The general solution can therefore be given as
\eqn\aio{
e^{-{2\ov n}\phi}=e^{-{2\ov n}\rho}=-\l^2 x^+x^- + {M\ov{\l}}
}
where $M$ is a constant.
\par
As shown in \faru\  
the solution \aio\ describes a black hole 
and the constant $M$ can be interpreted as the mass of the black hole.
The line element of the solution is
\eqn\liel{
ds^2=- {1\ov{({M\ov {\l}}-\l^2 x^+x^-)^n}}dx^+dx^-
}
and the scalar curvature $R$ is
\eqn\sca{
R=8e^{-2\rho}\dd_+\dd_-\rho = 4M\l n [{M\ov{\l}}-\l^2 x^+x^-]^{n-2}.
}
For later purposes it is convenient to recast the line element 
\liel\ in a chiral form
\eqn\chiral{
ds^2=-h(v,x)dv^2 +2dvdx.
}
To this end we perform the following coordinate transformation
\eqn\tra{
x^+=V, \ \ \ \ x^+ x^-={r\ov{\l}}
}
followed by
\eqn\trau{
V={1\ov{\l}}e^{\l v}, \ \ \ x={{({M\ov{\l}}-\l r)^{1-n}}\ov{2(1-n)\l}}
}
which yields the desired chiral form of the line element, namely
\eqn\kjus{
ds^2=-\{ 2 (1-n)\l x - {M\ov{\l}}[2 (1-n)\l x]^{{n\ov{n-1}}}\} dv^2
+ 2dvdx.
}
One should note that $x<0$ for $n>1$ and $x>0$ for $n<1$. 
\par
With this choice of coordinates
$v$ labels ingoing null lines and $x$ is an
appropriately normalized affine parameter on them.
\par 
As $h$ in \kjus\ does not depend on $v$, this metric has a Killing
field $\dd_v$ with Killing trajectories $x=\cst$
The function $h$ measures the norm of this Killing vector.
\par \n
In particular, the spacetime is stationary if the Killing field is
timelike ($h>0$) or homogeneous if the field is spacelike ($h<0$).
A zero of $h$ corresponds to a Killing horizon.
\par \n
From eq. \kjus\ our metric function $h$ is therefore
\eqn\acca{
h(x)=2(1-n)\l x- {M\ov{\l}}[2(1-n)\l x]^{{n\ov{n-1}}}
}
and for later use we give also the derivatives of $h$
\eqn\pri{
h^{'}(x)=2(1-n)\l \{ 1-{M\ov{\l}}({n\ov{n-1}})[2(1-n)\l x]^{{1\ov{n-1}}}\}
}
and 
\eqn\seco{
h^{''}(x)=-4M\l n [2(1-n)\l x]^{{{2-n}\ov{n-1}}}.
}
Finally in these chiral coordinates the scalar curvature $R$ is
simply 
\eqn\erre{
R=-h^{''}(x)=4M\l n [2(1-n)\l x]^{{{2-n}\ov{n-1}}}
}
and the dilaton field gives the coupling 
\eqn\dila{
e^{{2\ov n}\phi}=[2(1-n)\l x]^{{1\ov{n-1}}}
} 
and the cosmological constant term
\eqn\dilau{
e^{-2\phi}=[2(1-n)\l x]^{n\ov{1-n}}.
}

%%%%%%%%%%%%%%%%%%%%%%%%%%%%%%%%%%%%%%%%%%%%%%%
\newsec{ Space-time structure - The method}
%%%%%%%%%%%%%%%%%%%%%%%%%%%%%%%%%%%%%%%%%%%%%%%

In order to analyse the spacetime structure of our solutions and 
construct the corresponding maximally extended Penrose diagrams, 
we apply the method developed by Kl\"osch and Strobl in 
\klostr\
(to which we refer for the details).
\par \n
This method consists in a set of basic rules which allows, starting
from `building blocks', to construct the maximally extended Penrose
diagram corresponding to any metric function $h$ in eq. \chiral\
which depends on $x$ only (as in our case).
\par \n
The recipe is the following (for the case that the zeros of $h(x)$
are of odd order):
\par a) the number of zeros of $h$ determines the number of sectors
and their `orientation' in a fundamental building block.
$h>0$ corresponds to a stationary sector, $h<0$ to an homogeneous one.
For $n$ zeros of $h$ there are therefore $n+1$ sectors. A zero of $h$ 
corresponds to a Killing horizon which separates a stationary from
an homogeneous sector.
\par b) The end sector of the building block is a triangle if
\eqn\uau{
f(x)=\int^x {{du}\ov{h(u)}}
}
remains finite at the boundary; it is a square is $f(x)$ diverges.
\par c) Choose any sector and establish the symmetry axis, running
diagonally through the sector, transversal to the Killing line.
The `flip' symmetry
\eqn\flip{
x\to x, \ \ \ v\to -v +2f(x) 
}
amounts to a time reversal symmetry $t\to -t$ in the Schwarzschild
gauge
\eqn\schw{
ds^2 =- h(x)dt^2 + {{dx^2}\ov{h(x)}}.
}
This symmetry corresponds to a reflection symmetry of the sector 
with respect to the above symmetry axis.
Reflect the whole block at this symmetry axis and identify the 
corresponding sector.
\par d) Repeat this procedure for all sectors until one comes to an
end or ad infinitum.
\par e) If, after surrounding a vertex point, the sectors overlap,
identify the overlapping sectors such as to make a single sheet
around this vertex point.
\vskip .2in 
A boundary of the resulting diagram is null complete iff 
$x\to \pm \infty$ there. A boundary point is complete with respect to
all other geodesics iff
\eqn\time{
l(x)=\int ^x {{du}\ov{\sqrt{|c-h(u)|}}}
}
diverges there ($c$ is a constant).
\par
With these simple rules we proceed to the construction of the Penrose 
diagrams for the space-times of our theory.
The discussion will be divided in paragraphs, each corresponding to a 
particular range of the parameter $n$.

%%%%%%%%%%%%%%%%%%%%%%%%%%%%%%%%%%%%%%%%%%%%%%%%%%%%%%%%%%%%%%%%%%

\newsec{ Space-time structure - Examples }

%%%%%%%%%%%%%%%%%%%%%%%%%%%%%%%%%%%%%%%%%%%%%%%%%%%%%%%%%%%%%%%%%%%

%%%%%%%%%%%%%%%%%%%%%
\subsec{ a) $0<n<1$}
%%%%%%%%%%%%%%%%%%%%%
\par \n
In this case the range of variation of $x$ is $0<x<+\infty$.
At the boundary of this interval $h(0^+)\to -\infty$ and $h(+\infty)\to
2(1-n)\l x$.
\par \n
From the expression for the curvature scalar $R$ eq. \erre\ we see that
$x=0$ represents a singularity as $R\to +\infty$ there.
On the other hand $R(+\infty)\to 0$, so $x=+\infty$ represents the 
asymptotically flat region.
There is one horizon, the event horizon, located at 
\eqn\ura{
x_0=({{\l}\ov{M}})^{n-1}{1\ov{2(1-n)\l}},
}
where $h(x_0)=0$. This horizon divides 
the spacetime in a stationary region
for $x>x_0$ and an homogeneous one for $x<x_0$.
\par
The fundamental building block is given in Fig. 1a.
Note that the sector ending at the singularity $x=0$ is a triangle
( $f(0)$ is finite) and the singularity is spacelike.
The complete Penrose diagram is shown in Fig. 1b.
\par
Let us now consider the timelike geodesics equation in the Schwarzschild
gauge (eq. \schw)
\eqn\geo{
\dot x^2 = E^2 - h(x)
}
where a dot indicates proper time differentiation.
The qualitative behaviour of $h(x)$ is shown in Fig. 1c.
One notes that all geodesics fall into the singularity in a finite proper 
time 
\eqn\popi{
\tau=\int {{dx}\ov{\sqrt{E^2 - h(x)}}}.
}
The spacetime is therefore null and geodesically incomplete at $x=0$.
\par
Accelerated observers, moving on the trajectories $x=\cst >x_0$
see the geometry to be static and avoid to fall into the black
hole. The acceleration vector of these observers is
\eqn\acci{
a^{\alpha}=(0,{{h^{'}(x)}\ov{2}})
}
which is always directed to the right of the $x$ axis.
\par \n
Note that the ($x$,$t$) coordinates are not
inertial at infinity. The metric there can be approximated by
\eqn\mein{
ds^2=-2(1-n)\l x dt^2 + {{dx^2}\ov{2(1-n)\l x}}
}
which has not the Minkowskian form.
($x$, $t$) are, in fact, Rindler coordinates associated to an
acceleration $(1-n)\l$.
\par
By the transformation
\eqn\rind{
X=\sqrt{{{2x}\ov{(1-n)\l}}}\cosh (1-n)\l t, \ \ \ 
T=\sqrt{{{2x}\ov{(1-n)\l}}}\sinh (1-n)\l t
}
the metric acquires the usual Minkowski form
\eqn\mink{
ds^2=-dT^2 + dX^2.
}
The interesting property of this spacetime is the fact that, due to the 
asymptotic behaviour of $h(x)\sim 2(1-n)\l x$, all geodesic observers 
get captured by the black hole.
In fact, as one sees in Fig. 1c, outgoing (escaping) geodesics always
reach a turning point and then fall back into the black hole.
One can loosely think at this as 
 a consequence of the black hole being accelerated to the
right capturing all geodesic observers.
\par \n
Finally we mention the behaviour of the coupling $e^{{{2\phi}\ov n}}$, 
which can be seen to vanish asymptotically 
(weak coupling region) and diverge on 
the singularity (strong coupling region).

%%%%%%%%%%%%%%%%%%%%
\subsec{b) $1<n<2$}
%%%%%%%%%%%%%%%%%%%%

This case leads to a more complicated structure
and contains two completely different subcases.
\par \n
From the general expression eq. \trau\ we see that for $1<n<2$\ \ 
$x\in ]-\infty, 0]$. 
\par \n
$h(x)$ vanishes at $x=0$ and so does $R$.
$x=0$ corresponds, as we shall see,
 to an horizon, the acceleration horizon.
On the other boundary $R(-\infty)\to + \infty$, so $x=-\infty$
corresponds to a singularity of the spacetime.
\par \n
Besides $x=0$, $h(x)$ vanishes also at $x=x_0$, where $x_0$ is given as
before (see eq. \ura), but now $x_0$ is negative.
$x=x_0$ is the location of the black hole horizon and there $R(x_0) >0$.
\par \n
Finally $h^{'}(x)$ vanishes at 
\eqn\uau{
\tilde x = [{{(n-1)\l}\ov{nM}}]^{n-1} {1\ov{2(1-n)\l}}
}
and the behaviour of $h(x)$ is sketched in Fig. 2b.
\par \n
From these considerations one deduces the basic building block which is
represented in Fig. 2a.
\par \n
Note that the end sector is triangular, since $f(-\infty)$ is finite.
The spacetime is null complete at the boundary $x=-\infty$, but it is
timelike incomplete since the proper time $\tau(-\infty)$ is finite.
\par \n
Following similar arguments we find that the sector $x_0<x<0$ is a
square and that both timelike and null geodesics reach the boundary
$x=0$ at a finite value of their affine parameters.
The spacetime is regular at $x=0$; the metric there
has the same behaviour as in eq. \mein. 
Therefore $x=0$ represents an acceleration horizon with 
acceleration given as usual by its surface gravity $(n-1)\l$.
\par \n
It seems then natural to look for 
 analytical continuations of the
the metric across this horizon for positive
$x$ values.
\par \n
Of course, this doesn't mean that the solution is itself extensible
through $x=0$, since the dilaton field diverges there. However, we 
will relax any condition on $\phi$ and see how the function in eq. \acca\
can be extended beyond its original domain.
\par
Let us first discuss two smooth extensions which correspond to particular
values of the parameter $n$:
\par i) $(-1)^{{n\ov{n-1}}}=1$ even extension;
\par ii) $(-1)^{{n\ov{n-1}}}=-1$ odd extension.
\par
In the case i) the metric for $x>0$ has the following $h(x)$
\eqn\upi{
h(x)=2(1-n)\l x - {M\ov{\l}} [2(n-1)\l x]^{n\ov{n-1}},
}
whereas for the case ii) we have
\eqn\icci{
h(x)= 2(1-n)\l x + {M\ov{\l}} [2(n-1)\l x]^{n\ov{n-1}}.
}
We shall consider first the case i).
\par \n
Being $h(x)$ defined by eq. \upi, we have that $h(x)$ in the interval
$x\in [0, +\infty[$ vanishes only at $x=0$ and it is negative everywhere
else (within this sector the spacetime is homogeneous).
$h^{'}(x)$ is always negative and 
\eqn\era{
R=-h^{''}(x)=4\l n M [2(n-1)\l x]^{{{2-n}\ov{n-1}}} .
}
From this equation we see that $x=+\infty$ corresponds to a singularity
and, being $f(+\infty)$ finite, the sector $0<x<+\infty$ is a triangle.
\par \n
Adding together all information we can infer the behaviour of $h(x)$ for
$x\in ]-\infty, +\infty[$ which is depicted in Fig. 3c.
Also given are the building block and the complete Penrose diagram
in Figs 3a and 3b.
\par \n
We see that there are two horizons in the spacetime, $x=x_0$ and $x=0$.
All timelike geodesics, but one, start and end at the spacelike
singularities ($x=\pm\infty$) reaching them in a finite proper time.
The spacetime is timelike incomplete at the boundaries, but null complete.
\par \n
The only geodesic which avoids falling into the singularity is $x=\tilde x$.
Also accelerated observers $x=\cst=k$ ($x_0<k<0$) escape the singularities.
Their acceleration 
is directed to the right for $x<\tilde x$ and to the
left for $\tilde x <x <0$.
\par \n
$x=0$ is the acceleration horizon for these comoving Rindler observers.
Starting and end points of these trajectories (and of the $x=\tilde x$ 
geodesic) are the dark dots in the Penrose diagram of Fig. 3b, which are 
the only points at infinity of this spacetime.
No stationary motion is possible for $x<x_0$ and $x>0$.
\par
The behaviour of the dilaton field for $x>0$ can be seen from 
\dila\ and \dilau.
It is
\eqn\coco{
e^{{2\ov n}\phi}=-[2(n-1)\l x]^{{1\ov{n-1}}},
\ \ \ e^{-2\phi}=[2(n-1)\l x]^{{n\ov{1-n}}}.
}
Thus, while for $x<0$ $\phi$ is real and the two terms in \coco\ are
positive, for $x>0$ $\phi$ becomes complex and acquires a constant imaginary 
part \foot{ If one imposes a reality condition on $\phi$ (see for example
\mignemi) 
the surface
$x=0$ must be considered a boundary of the spacetime, which is then
timelike and null incomplete there.}
\eqn\imi{
\phi \to  \phi +i{{\pi n }\ov{2(n-1)}}.
} 
This explains why $e^{{2\ov n}\phi}$ becomes negative
while $e^{-2\phi}$ stays $>0$. The coupling $e^{{2\ov n}\phi}$ 
vanishes for $x=0$ and 
diverges on the singularity located at $x=+\infty$.
\par 
Now consider the case ii), with $h(x)$ given by eq. \icci. 
\par \n
$h(x)$ vanishes, for $x\in [0,+\infty[$, not only at $x=0$ but also at
$x=-x_0$, which is the location of another horizon, the naked singularity
horizon. 
The curvature scalar $R$ vanishes at $x=0$ and is negative in the interval
$0<x<+\infty$. In particular $R(+\infty)=-\infty$, signalling the fact 
that $x=+\infty$ is a singularity. Being $h(x)>0$ for $x\to +\infty$
this singularity is timelike unlike the previous one. 
\par \n
Examination of $f(x)$ reveals that the sector $0<x<-x_0$ is squared, but
the last sector $x>-x_0$ is a triangle.
The corresponding building block is given in Fig. 4a, from which one deduces
as resulting Penrose diagram of the maximal analytical extension of
the spacetime a complicated network structure reproduced in Fig. 4b.
\par 
The spacetime is again null complete, but timelike incomplete at the
boundaries 
\par \n 
$x=\pm \infty$.
The graph reported in Fig. 4c shows the behaviour of $h(x)$, $x\in ]-\infty,
+\infty[$.
\par \n
We see that $h(x)$ vanishes at three horizons, $x=\pm x_0$ and $x=0$.
$x=x_0$ is the black hole horizon, $x=-x_0$ the naked singularity horizon
and $x=0$ the acceleration horizon where the spacetime becomes flat
(again in Rindler like coordinates).
\par \n
The asymptotic behaviour of $h(x)$ for $x\to +\infty$ implies that no
timelike geodesic reaches the timelike singularity. They all bounce back.
There is bounded geodesic motion for $x>\tilde x$, with turning points
at $x>-x_0$. All the other geodesics get eventually captured by the
black hole singularity at $x=-\infty$. As before accelerated observers
(and the $x=\tilde x$ geodesic observer) avoid falling into the singularity
and reach the asymptotic points of the diagram.
\par \n
The acceleration of the $x=\cst=k$
($x_0<k<0$, $k>-x_0$)  observer is directed to the right for $x<\tilde x$
and $x>-x_0$ and to the left for $\tilde x < x < 0$.
\par
Finally the dilaton, when $x>0$, takes an imaginary part as in eq. \imi,
but now
\eqn\gigi{
e^{{2\ov n}\phi}=[2(n-1)\l x]^{{1\ov{n-1}}},\ \ \ 
e^{-2\phi}=-[2(n-1)\l x]^{{n\ov{1-n}}}.
}
\par
For values of $n$ different from the ones considered in the cases i) and ii)
one can extend the metric to positive values of $x$ introducing an
absolute value, namely defining
\eqn\oio{
h(x)=2(1-n)\l x - {M \ov{\l}} [2\l |x(1-n)|]^{{n\ov{n-1}}}.
}
The resulting spacetime has a structure described by a Penrose diagram 
which is the same as the one given in Fig. 3b for the case i).
\par \n
Alternatively, one could also consider
\eqn\cupo{
h(x)=2(1-n)\l x \{ 1 - {M\ov{\l}} [2\l |(1-n)x| ]^{{1\ov{n-1}}}\} .
}
The spacetime which results from this choice of $h(x)$ is described by the
Penrose diagram of Fig. 4b as for the case ii).
\par \n
In these last two cases the coupling $e^{{2\ov n}\phi}$ and the cosmological
constant term $e^{-2\phi}$ behave correspondingly as in i) and ii).
\par \n
The same type of extensions have been considered in Ref \trivedi\ across
the horizon of a semiclassical extremal black hole.

\par 
Note, finally, that due to the lack of a `true' asymptotic region 
in the diagrams of Figs 3b and 4b, $x=x_0$ cannot be
considered an event horizon for our spacetimes.

%%%%%%%%%%%%%%%%%%%%%%%%%%%%%%%%%%%%%%%%%%
\subsec{c) $n=2$}
%%%%%%%%%%%%%%%%%%%%%%%%%%%%%%%%%%%%%%%%%%

This case merits a discussion for himself because the resulting spacetime
has constant curvature.
In this case
\eqn\guru{
h(x)=-2\l x - 4\l M x^2.
}
$h(x)$ vanishes for $x=0$ and $x_0=-{1\ov{2M}}$. 
$h^{'}(x)$ vanishes at $\tilde x=-{1\ov{4M}}$ and is positive 
for $x<\tilde x$ and negative elsewhere. The behaviour of $h(x)$ is depicted
in Fig. 5c.
\par \n
The extension of $h(x)$ across $x=0$ is again considered without any
requirement on $\phi$.
\par \n
The scalar curvature $R$ is a positive constant
\eqn\desi{
R=8\l M
}
and the spacetime is everywhere regular. 
The sectors $-\infty <x<x_0$ and $x>0$ are triangular, whereas $x_0<x<0$
is a square. 
\par \n
The fundamental building block and the
Penrose diagram are depicted in Figs 5a and 5b.
\par \n
The resulting spacetime is both null and timelike complete at the
boundaries $x=\pm \infty$.
There are two horizons at $x=0$ and $x=x_0$. Accelerated observers at
$x=\cst =k$ ($x_0<k<0$) have their acceleration directed to the right for
$x<-{1\ov{4M}}$ and to the left for $x>-{1\ov{4M}}$. For $x=-{1\ov{4M}}$
the trajectory is a geodesic.
Typically geodesics run from one asymptotic region to the other or back.
\par
A final comment concerns the dilaton field and the two terms
\eqn\ciap{
e^{{2\ov n}\phi}=e^{\phi}=-2\l x,\ \ \ e^{-2\phi}=(-2\l x)^{-2}.
}
$\phi$ acquires, according to \imi, an imaginary part $+i\pi$.
$x=0$ is a weak-coupling region, while $x=\pm \infty$ are strong-coupling.

%%%%%%%%%%%%%%%%%%%%%%%%%%%%%%%
\subsec{d) $n>2$}
%%%%%%%%%%%%%%%%%%%%%%%%%%%%%%%%

Let us finally discuss briefly the case $n>2$. 
\par \n
Being $x<0$, we see that
$h(x)$ vanishes at $x=0$ and $x=x_0$ ($<0$).
The only horizon is at $x=x_0$. The null surface $x=0$ is a singularity
because the curvature diverges there, $R(0)\to + \infty$.
\par
The behaviour of $h(x)$ is sketched in Fig. 6c.
One deduces that the sector $x_0 < x <0$ is squared, whereas the 
sector $x<x_0$ is triangular since $f(-\infty)$ is finite.
\par \n
The fundamental building block is depicted in Fig. 6a and the
resulting Penrose diagram is given in Fig. 6b.
Since null and timelike geodesics reach $x=0$ for finite values of their
affine parameters, the spacetime is null and timelike incomplete at this
boundary.
\par \n
Stationary motion exists only for $x>x_0$. $x=\cst =k$  ($x_0<k$)
observers have their acceleration directed to the right (increasing $x$
values, towards the singularity) 
for $x<\tilde x$ and to the left 
for $x>\tilde x$.
Timelike geodesics coming from past infinity either have a turning point
and continue to future infinity or proceed towards the singularity.
\par
Note finally that the coupling $e^{{2\ov n}\phi}$ 
 vanishes on the singularity (which appears therefore in the 
weak coupling regime) and diverges asymptotically.  
\par
Although diverging at the singularity,
the curvature $R$ is integrable and the singularity is in some sense
`mild' as the tidial distortion remains finite when crossing the 
 surface $x=0$.
\par \n
One might extend the metric across $x=0$ and obtain a spacetime whose 
Penrose diagram is similar to the one obtained in the case
$n=2$ (see Fig. 5b).

%%%%%%%%%%%%%%%%%%%%%%%%%%%%%%%%%%%%%%%%%%%%%%%%%%%%%%%%%%%%%%%%%%%%%%%%%%%

\newsec{ Conclusions}

%%%%%%%%%%%%%%%%%%%%%%%%%%%%%%%%%%%%%%%%%%%%%%%%%%%%%%%%%%%%%%%%%%%%%%%%%%%

In this paper we have discussed the maximal extension of the space-times 
described by the solution of eqs \eqmo\ and \dquattro.
\par
In the case $n=1$ eq. \liel\ reduces to the usual black
hole solution of Ref \cghs, for which 
there exists an asymptotic minkowskian and static frame. 
This is not true anymore for $n\neq 1$ and this simple fact
gives the interesting implications that emerged 
in section 4, which we here summarize.
\par
For $0<n<1$, as for $n=1$, the Penrose diagram is the same as that of
a Schwarzschild black hole. However, the static frame ($x$, $t$) of eq. \schw\ 
is not asymptotically minkowskian,
but Rindler like (see eq. \mein). 
One can introduce the inertial frame ($X$, $T$), defined in \rind,
but as a result the metric is nonstatic.
\par \n
As a consequence of the fact that the 
black hole can be thought to be accelerated, 
 all geodesic observers get soon or later captured and inevitably
end their journey into the singularity. The only way to avoid this fate
and reach safely the asymptotic region
along a timelike trajectory is to have a nonzero acceleration, as for
instance the curves $x=\cst$
\par
For values of $n$ such that $1<n<2$ a completely different causal structure
emerges, as it is shown in Figs 3b and 4b.
\par \n
Fig. 3b suggests the existence of two black holes, one at each end 
of the $x$ axis. They are both attracting and the combined action
of their gravitational fields pulls any observer at $x<\tilde x$
(see eq. \uau) 
towards the black hole located at $x=-\infty$ and those at $x>\tilde x$
towards the other one. At $x=\tilde x$,
although the scalar curvature $R$ doesn't vanish,
there is an equilibrium between the two forces.
\par \n
$R$ vanishes at $x=0$, the location of the acceleration horizon for
Rindler observers. Inertial observers, on the other hand, can cross
this horizon and enter a 
collapsing cosmological region defined for $x>0$, as in the case 
of a black hole immersed in a closed universe.
\par \n
The diagram of Fig. 4b has qualitatively a similar description till
 $x=0$, but now the gravitational collapse of the $x>0$ region 
 is not strong enough to cause the `big crunch' singularity.
Rather a wormhole appears, through which one can safely travel
to a new
universe similar to the one just left. 
This multi-black hole structure possesses both spacelike and timelike
singularities, all strong-coupling in the dilaton field.  
\par 
In the case $n=2$ the structure of the Penrose diagram of Fig. 5b is 
similar to
that of Fig. 3b except for the important fact that the singularities are
replaced by asymptotic well-behaved regions of the two-dimensional metric.
\par
Finally, for $n>2$ the form of the diagram is the same as in Fig. 1b,
but with the singularity and the asymptotic regions exchanged: the singularity
is now light-like.
\par 
The spacetime features we have explored here are not peculiar of our
two-dimensional theory eq. \accla.
They appear also in the context of
 higher-derivative $2d$ gravity, as for example in \schmidt.   
\par \n
Let us consider the action 
\eqn\deriv{
S=\int d^2 x \sqrt{-g} [ {{R^{k+1}}\ov{16}} + \Lambda ],
}
where $k$ is a positive integer.
The trace and the traceless part of the field equations are
\eqn\suomi{
{k\ov{16}} R^{k+1} + {{(k+1)}\ov{16}}\nabla^2 R^k = \Lambda ,\ \ \
\nabla_{\mu} \dd_{\nu} R^k - {1\ov 2} g_{\mu\nu}\nabla^2 R^k =0.
}
In the Schwarzschild gauge eq. \schw, where $R=- h^{''}(x)$,
eqs \suomi\ give the third-order equation  
\eqn\perni{
{{k(k+1)}\ov{16}} h^{'''}(x)h^{'}(x)+ 
\Lambda [-h^{''}(x)]^{1-k}={{k [h^{''}(x)]^2}\ov{16}}.
}
The general solution of eq. \perni\ is
\eqn\cupo{
h(x)=C+B\Lambda  x +D x^{2+{1\ov k}},
} 
where
\eqn\uppi{
B=-{{- 16[-D (2+ {1\ov k})(1+ {1\ov k}]^{1-k}}\ov{k(1+ {1\ov k})^2(2+{1\ov
k})D}}.
}  
The scalar curvature of these solutions is
\eqn\acco{
R=-(2+{1\ov k})(1+{1\ov k})D x^{{1\ov k}}.
}
Note that for $C=0$ $h(x)$ has the same structure of our eq. \acca\
with $k={{n-1}\ov{2-n}}$. 
\par \n
One can proceed with the analysis of the spacetime structure of these
solutions along the line of the previous paragraph.
Here we just mention that
for $k=1$, $C=0$ and $D={2\ov 3}$, as considered in Ref \klostr,  
the function $h(x)$ has the same behaviour of eq. \acca\ with $n={3\ov 2}$
 and the resulting Penrose diagram is represented in Fig. 4b. 
\vskip .1in
 {\it Note added:} After completion of this work we learnt that
the spacetime structure of our model has been briefly discussed 
in a recent preprint by M.O. Katanaev, W. Kummer and H. Liebl,
{\it On the Completeness of the Black Hole Singularity in 2d Dilaton 
Theories}, gr-qc/9602040. These authors seem, however, not to have paid
much attention to our model because of ``problems of interpretation" 
related to the asymptotic Rindler structure of the spacetime.
%%%%%%%%%%%%%%%%%%%%%%%%%%%%

\bigskip\bigskip

%%%%%%%%%%%%%%%%%%%%%%%%%%%%%%%%%%%%%%%%%%%%%%%%%%%%%%%%%%%%%%%%%%%%
\noindent $\underline{\rm Acknowledgements}$:
%%%%%%%%%%%%%%%%%%%%%%%%%%%%%%%%%%%%%%%%%%%%%%%%%%%%%%%%%%%%%%%%%%%%
 We would like to thank D. Amati and J.G. Russo for useful discussions.

  \listrefs
\vfill\eject
%\vskip .5in
{
 \epsfxsize=6.5cm \epsfysize=9.0cm 
 \epsfbox{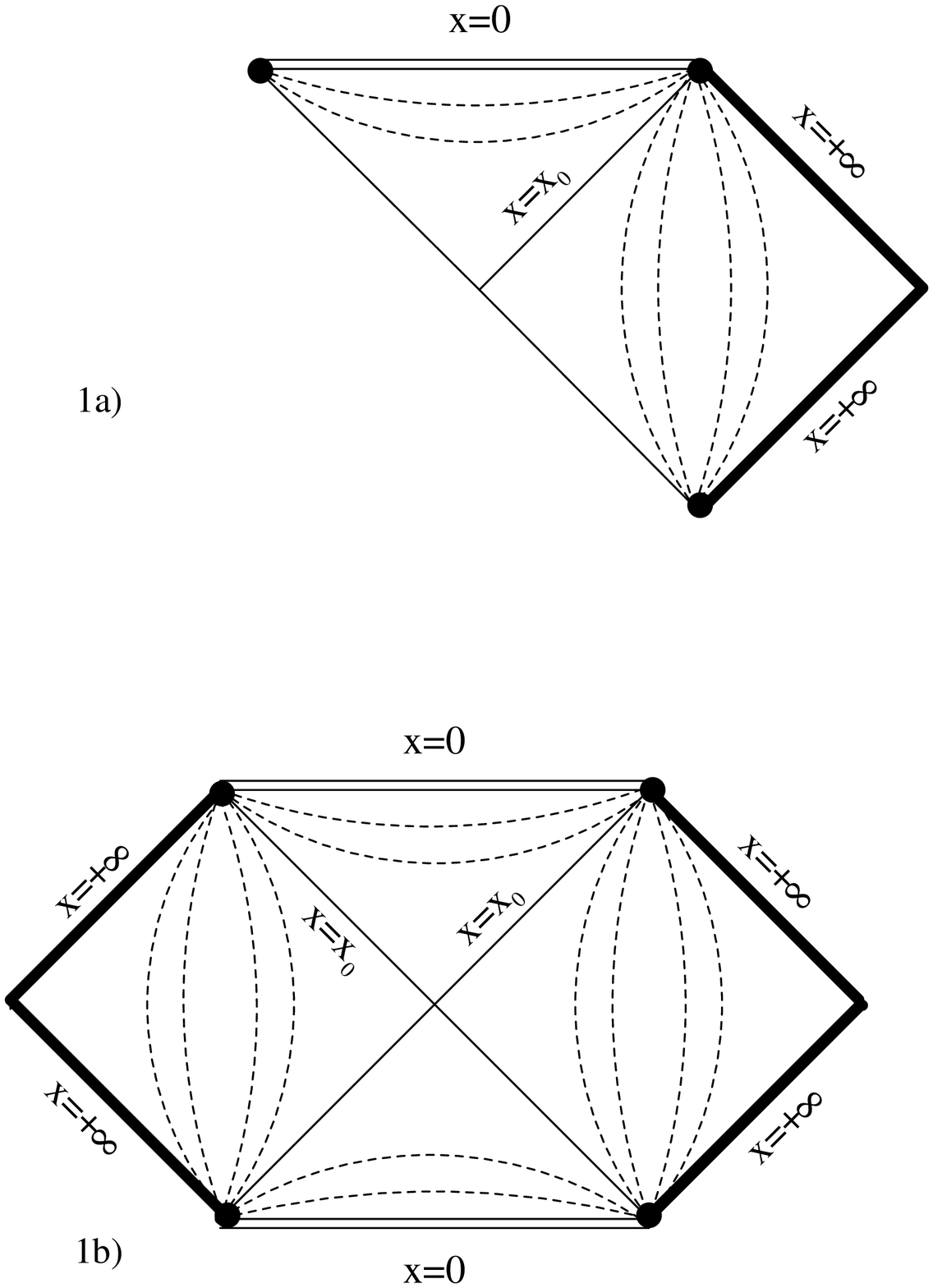}
 }
{\bf {Fig. 1:}}{
Fundamental building block in a) and Penrose diagram in b) for the case
$0<n<1$. 
Double lines represent the singularity, 
dashed lines the curves $x=\cst$, regular lines the horizons and 
thick lines the asymptotic region.}

{
\epsfxsize=6.5cm \epsfysize=7.0cm 
\epsfbox{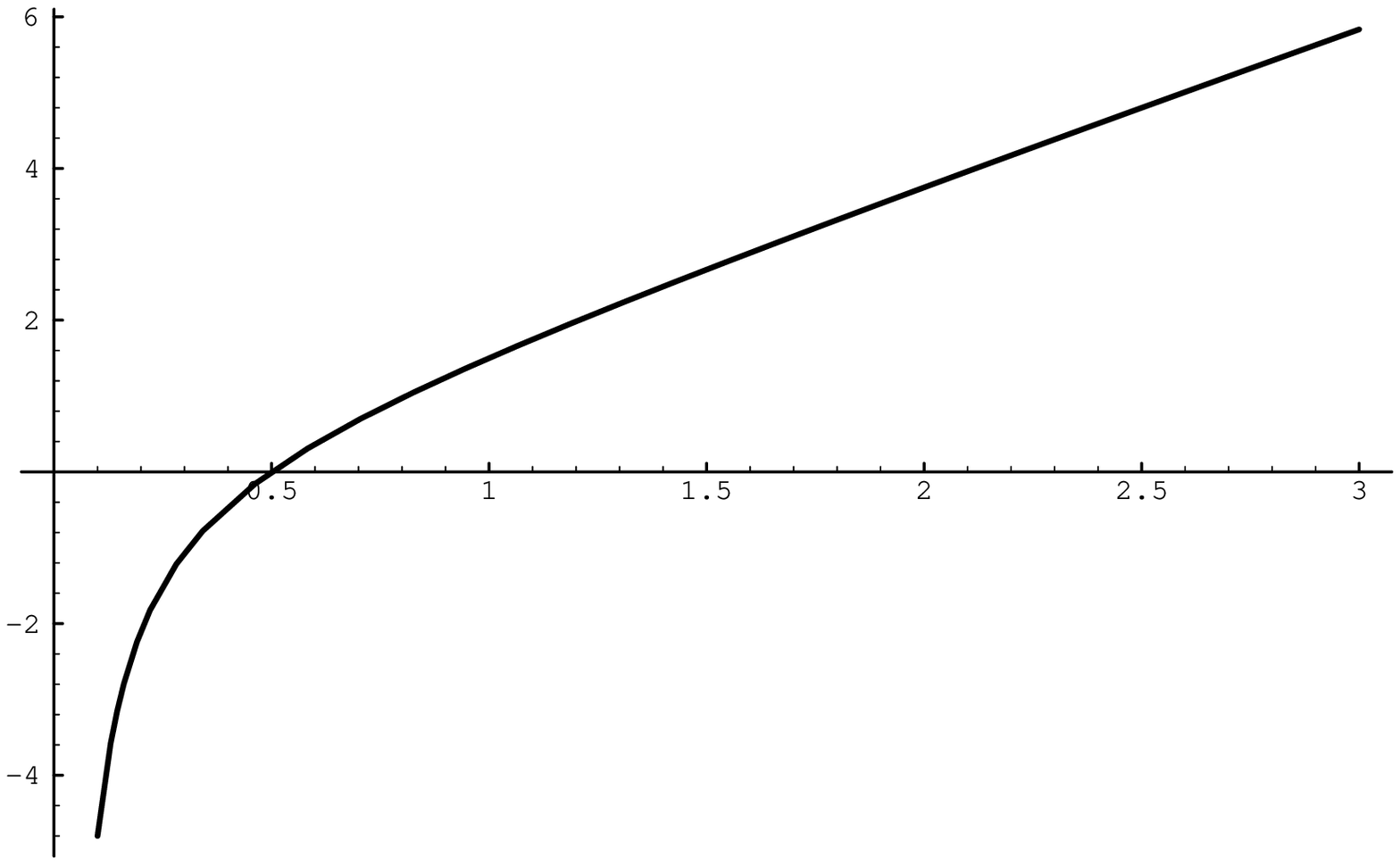}}

{\bf {Fig. 1c:}}{ 
Graph of $h(x)$ for $0<n<1$ (here $\l =M={1\ov{n-1}}$).}

\vfill\eject

%\vskip .5in
{
\epsfxsize=8cm \epsfysize=5.5cm 
\epsfbox{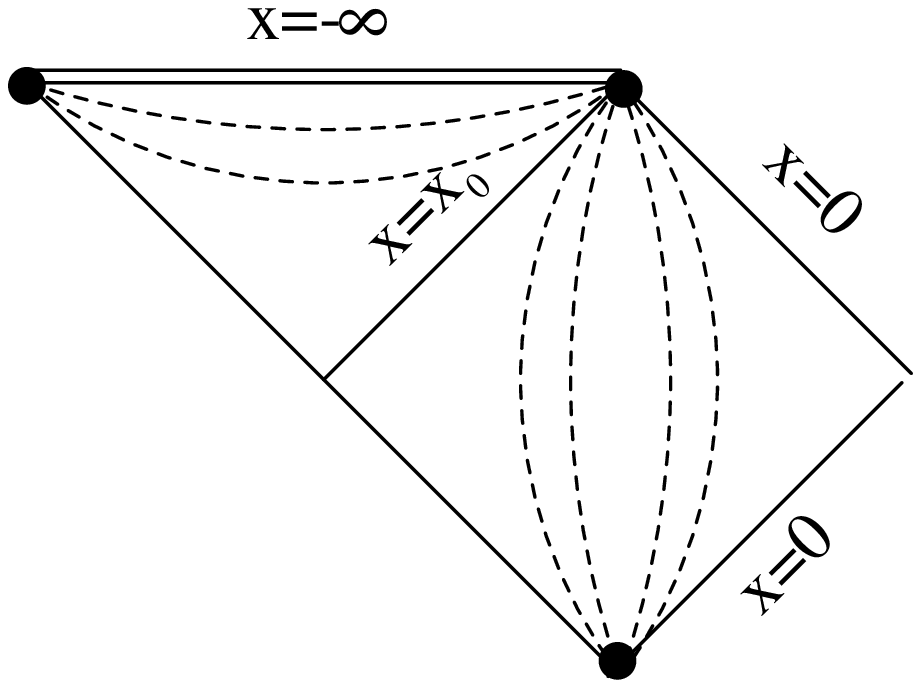}}
{\bf {Fig. 2a:}}{
Building block for the case $1<n<2$ till $x=0$.}

{
\epsfxsize=7.5cm \epsfysize=8.5cm
\epsfbox{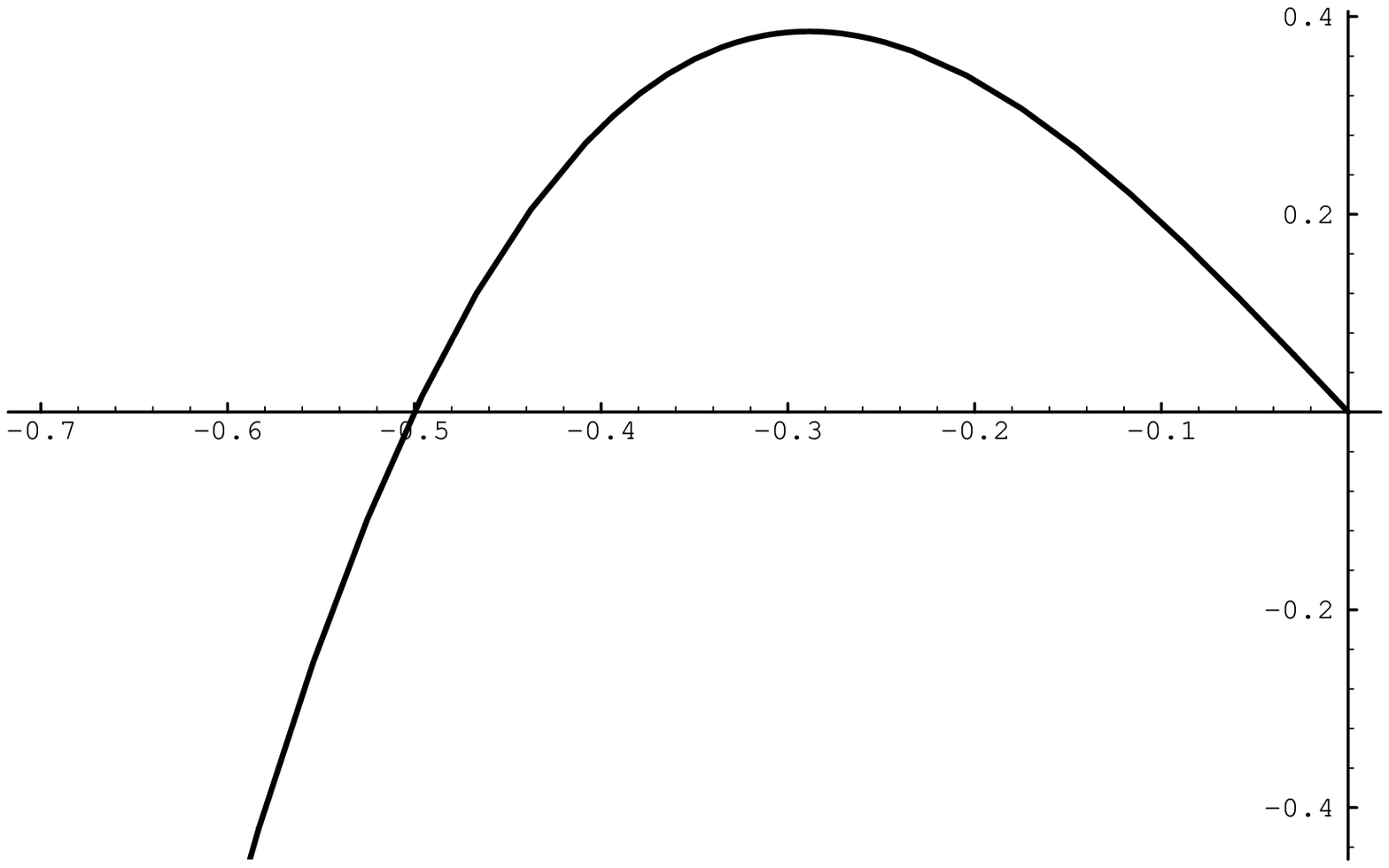}}
{\bf {Fig. 2b:}}{
Graph of $h(x)$ for the case in Fig. 2a.}
\vfill \eject

\vfill\eject
%\vskip -20 true pt

{
 \epsfxsize=9,5cm \epsfysize=8.5cm 
 \epsfbox{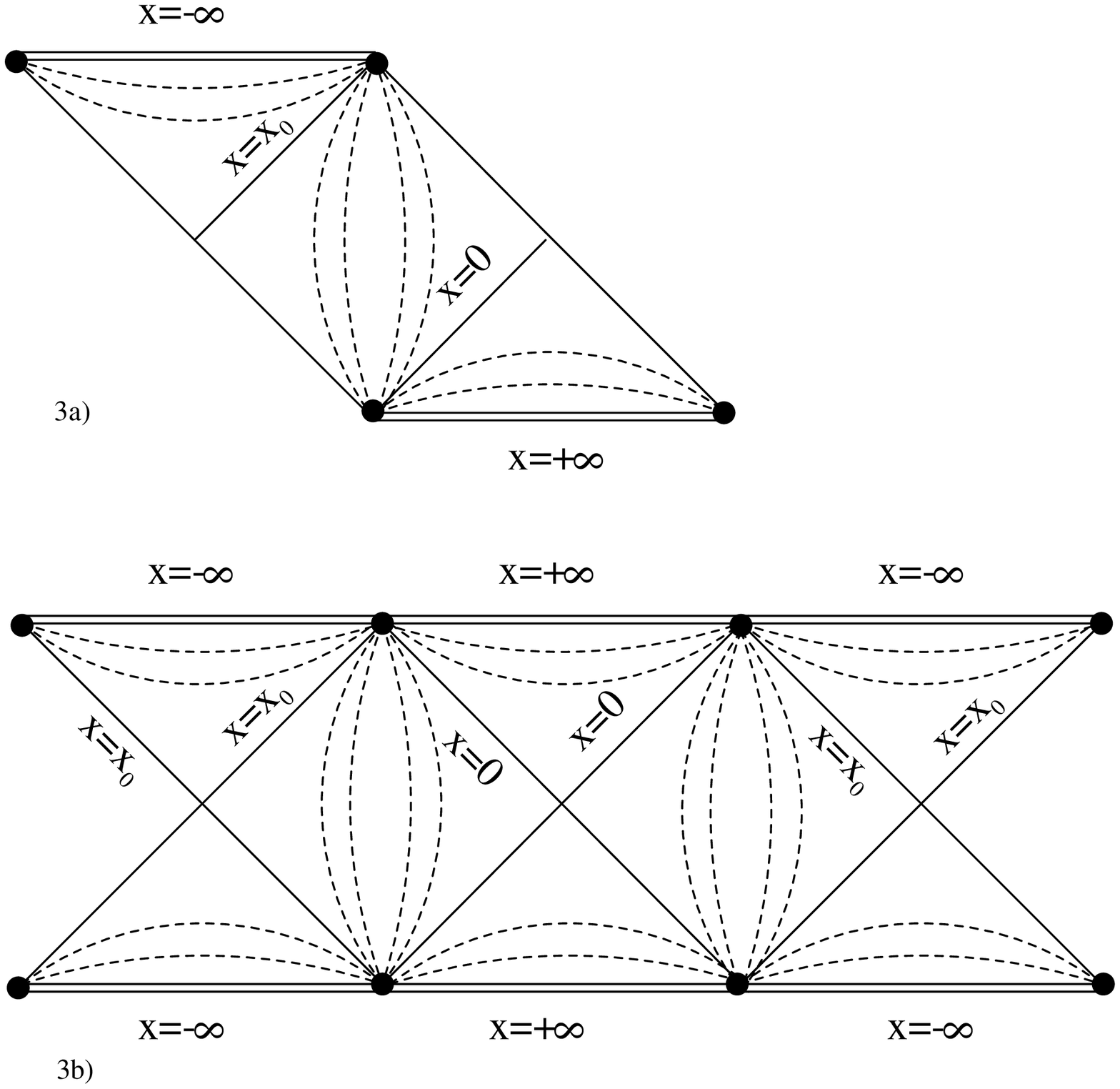}}
{\bf{Fig. 3:}}{
Fundamental building block in a) and Penrose diagram in b) for the
extension i).}

{
\epsfxsize=7,5cm \epsfysize=8cm 
\epsfbox{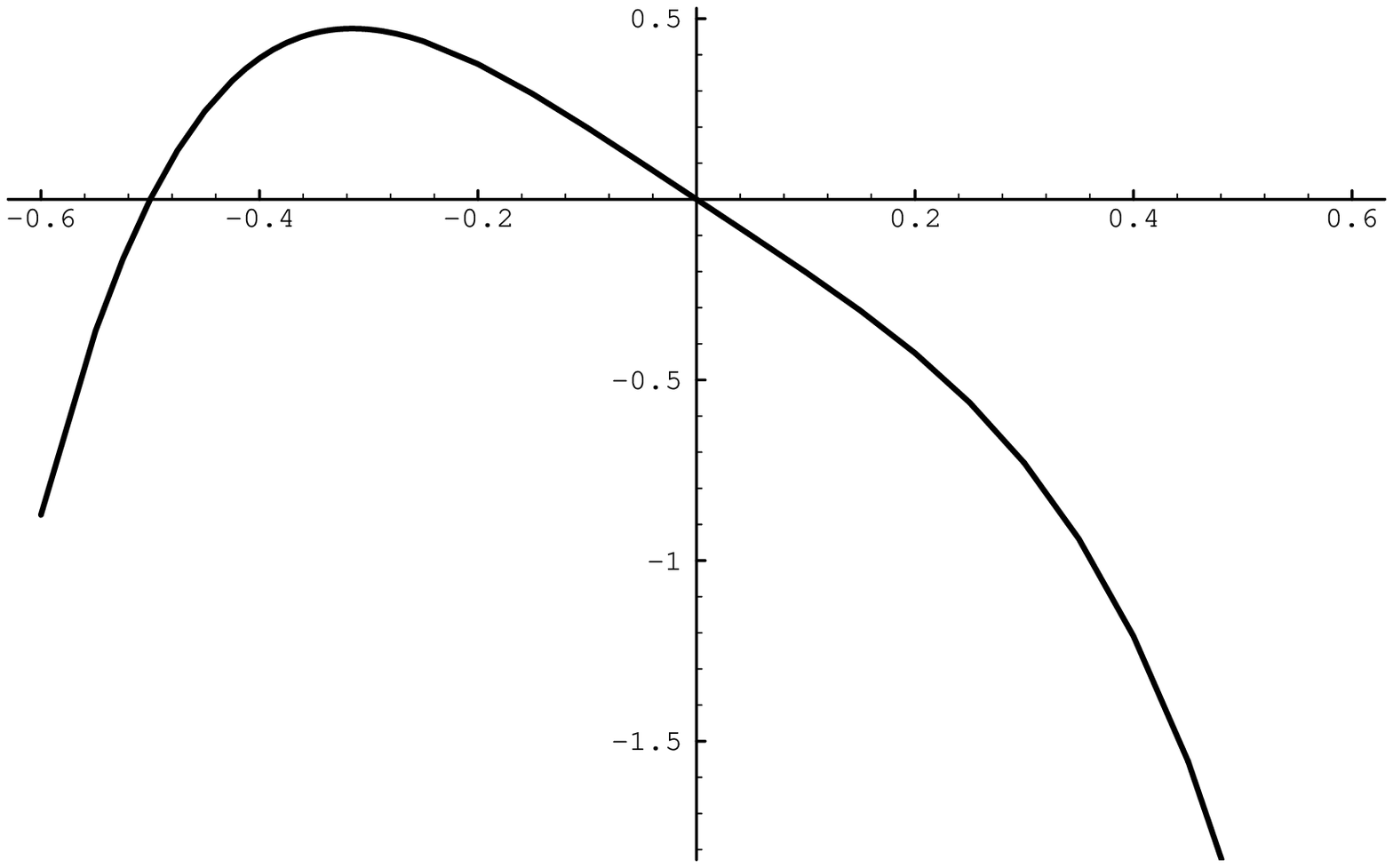}}
{\bf{Fig. 3c:}}{
Graph of $h(x)$ in the case i).}
\vfill\eject
%\vskip -20 true pt

{
 \epsfxsize=7cm \epsfysize=10cm 
 \epsfbox{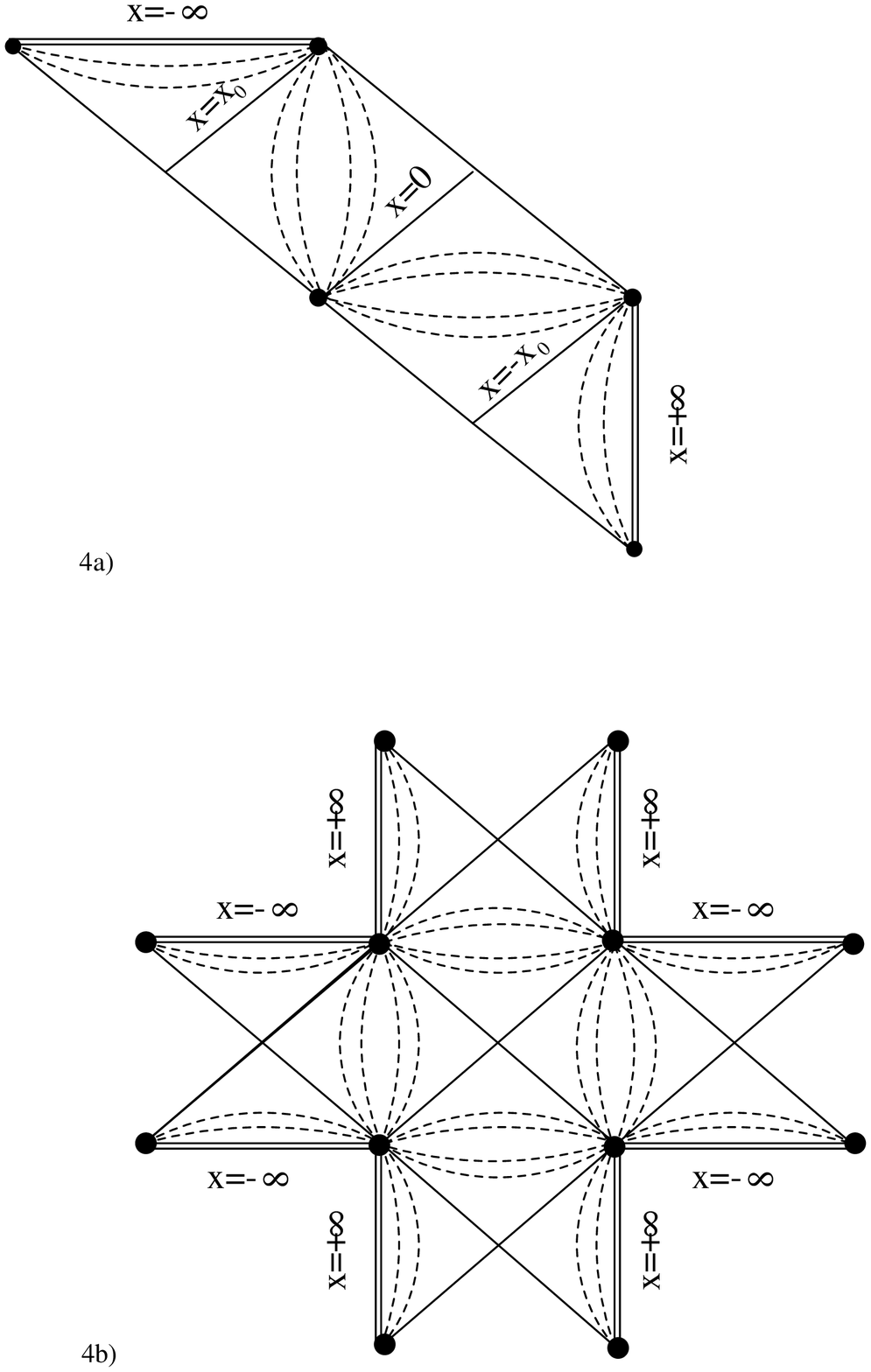}}
{\bf{Fig. 4:}}{
Fundamental building block in a) and Penrose diagram in b) for the extension
ii).}

{
\epsfxsize=7.5cm \epsfysize=7.0cm 
\epsfbox{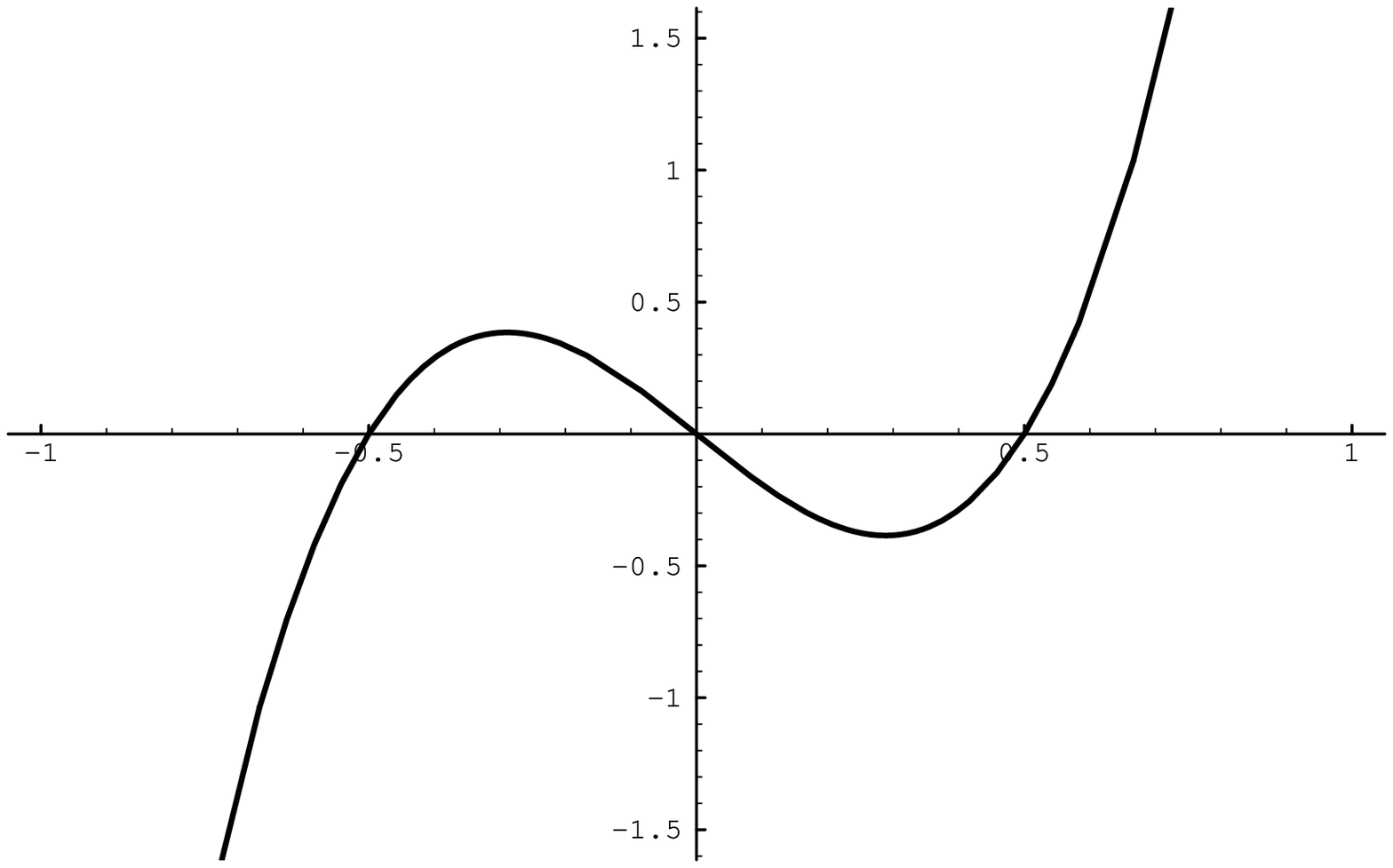}}
{\bf{Fig. 4c:}}{
Graph of $h(x)$ in the case ii).}
\vfill\eject
%\vskip -20 true pt

{
 \epsfxsize=10cm \epsfysize=9cm 
 \epsfbox{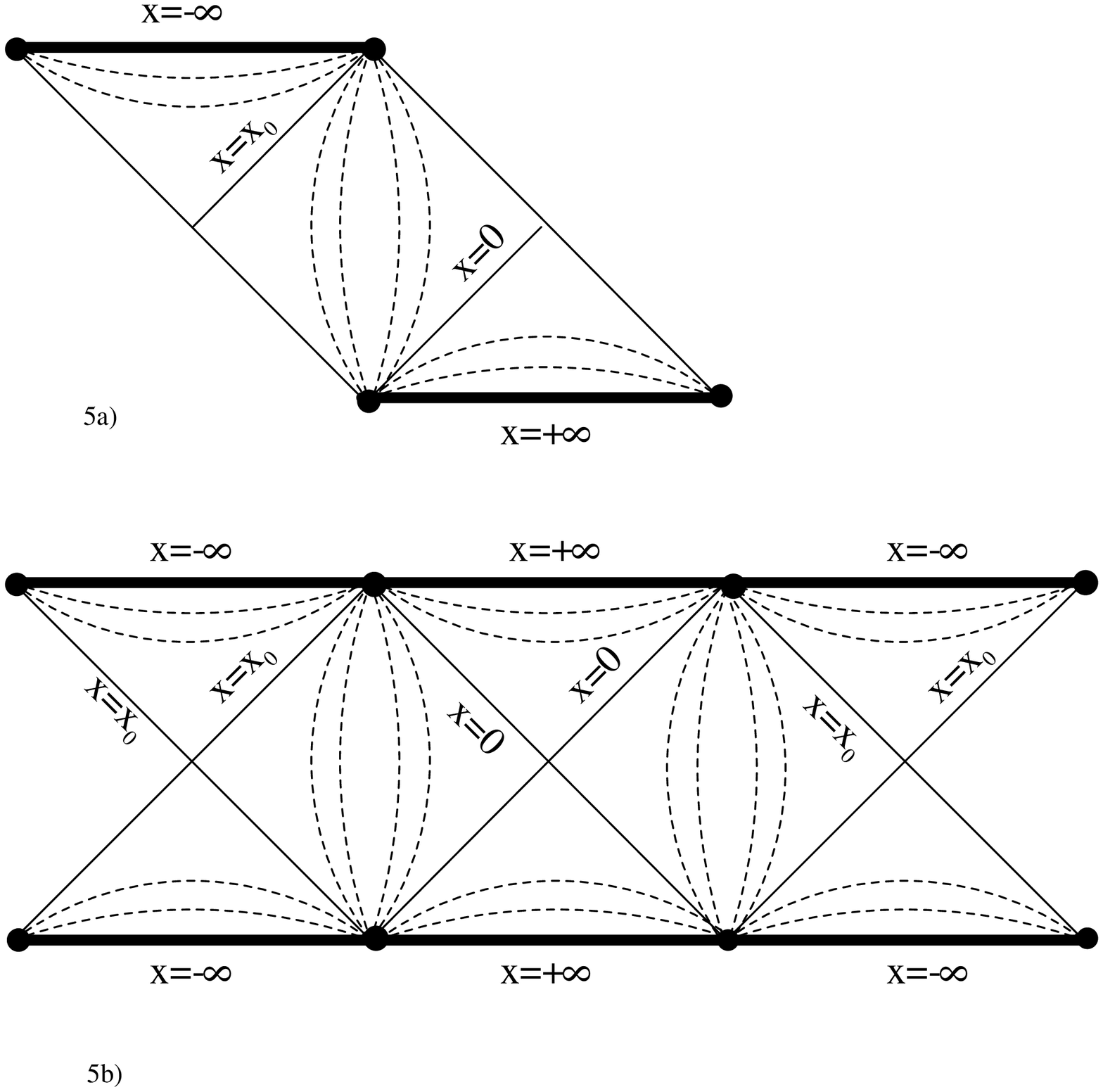}}
{\bf{Fig. 5:}}{
Fundamental building block in a) and Penrose diagram in b) in the case $n=2$.}

{
\epsfxsize=7cm \epsfysize=7.5cm 
\epsfbox{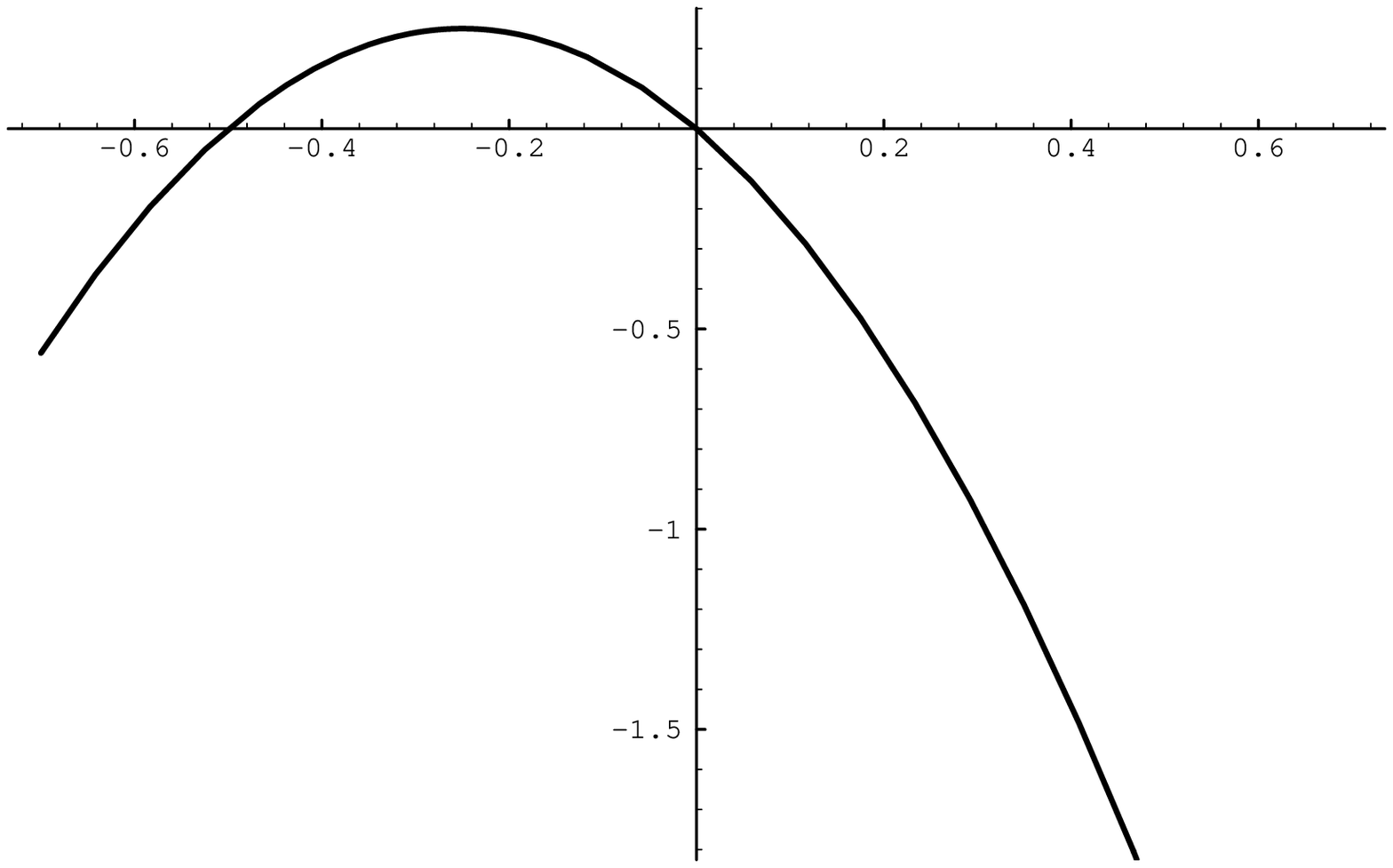}}
{\bf{Fig. 5c:}}{
Graph of $h(x)$ for $n=2$.}
\vfill\eject
%\vskip -20 true pt

{
 \epsfxsize=7cm \epsfysize=10cm 
 \epsfbox{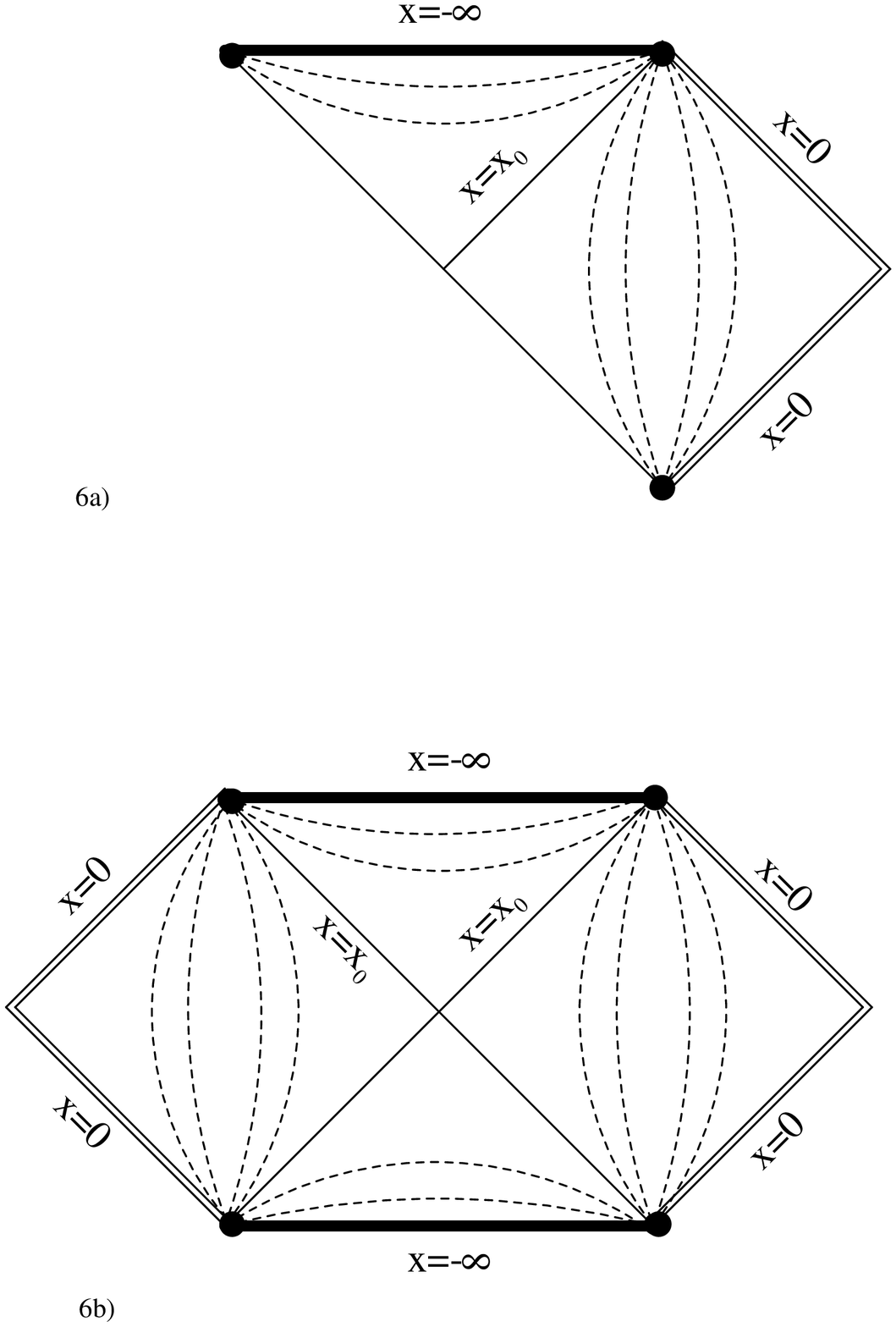}}
{\bf{Fig. 6:}}{
Fundamental building block in a) and Penrose diagram in b) in the case $n>2$.}

{
\epsfxsize=7cm \epsfysize=7cm 
\epsfbox{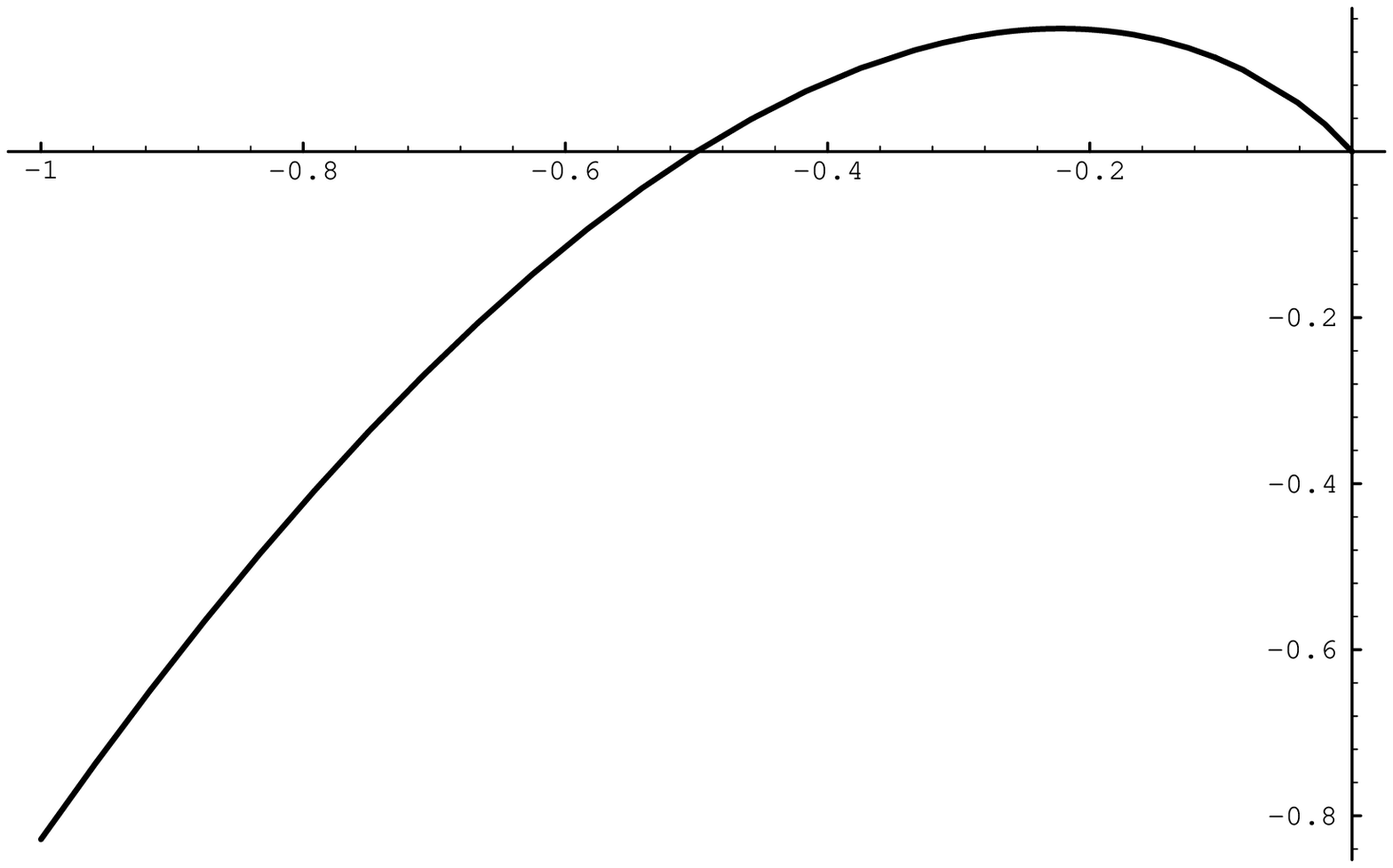}}
{\bf{Fig. 6c:}}{
Graph of $h(x)$ for $n>2$.}
\vfill\eject
\end